\documentclass[aps,prb,10pt,twocolumn,superscriptaddress,showpacs]{revtex4-2}

\usepackage[font={small}, justification=justified,singlelinecheck=false]{caption}
\usepackage[justification=justified,singlelinecheck=false]{subcaption}
\usepackage{soul, color}
\usepackage{scalerel}
\usepackage{physics}
\usepackage{float}
\usepackage{graphicx}
\usepackage{dcolumn}
\usepackage{bm}
\usepackage{hyperref}
\hypersetup{colorlinks, linkcolor={red}, citecolor = {blue}, urlcolor={blue}}

\newcommand {\hide} [1] {}
\usepackage{comment}

\let\oldref\ref
\renewcommand{\ref}[1]{(\oldref{#1})}

\begin{document}

\title{Bang-bang algorithms for quantum many-body ground states:\\
a tensor network exploration}

\author{Ruoshui Wang}
\email{rw552@cornell.edu}
\affiliation{Cornell University, Ithaca, New York, 14853 USA}

\author{Timothy H. Hsieh}
\affiliation{Perimeter Institute for Theoretical Physics, Waterloo, Ontario, N2L 2Y5 Canada}
\affiliation{University of Waterloo, Waterloo ON, N2L 3G1, Canada}

\author{Guifr\'e Vidal}
\affiliation{Google Quantum AI, Mountain View, CA 94043, USA}

\date{\today}

\begin{abstract}
We use matrix product techniques to investigate the performance of two algorithms for obtaining the ground state of a quantum many-body Hamiltonian $H = H_A + H_B$ in infinite systems. 
The first algorithm is a generalization of the quantum approximate optimization algorithm (QAOA) and uses a quantum computer to evolve an initial product state into an approximation of the ground state of $H$, by alternating between $H_A$ and $H_B$. We show for the 1D quantum Ising model that the accuracy in representing a gapped ground state improves exponentially with the number of alternations.  The second algorithm is the variational imaginary time ansatz (VITA), which uses a classical computer to simulate the ground state via alternating imaginary time steps with $H_A$ and $H_B$.  We find for the 1D quantum Ising model that an accurate approximation to the ground state is obtained with a total imaginary time $\tau$ that grows only logarithmically with the inverse energy gap $1/\Delta$ of $H$.  This is much faster than imaginary time evolution by $H$, which would require $\tau \sim 1/\Delta$. 
\end{abstract}

\pacs{Valid PACS appear here}
\maketitle

\section{\label{sec:level1}Introduction}

Characterizing the emergent collective properties of quantum many-body systems is among the most challenging problems in theoretical physics. Such problems are essential in understanding many strongly correlated materials in condensed matter physics or complex molecules in quantum chemistry. For these problems, it is often important to understand the properties of the ground state of a quantum many-body Hamiltonian. However, an exact characterization of the ground state is generically precluded by the exponentially large dimension of the system's Hilbert space.
There have been numerous computational approaches, ranging from large-scale exact diagonalization to Monte Carlo and tensor network methods, which have made substantial progress on many frontiers.  However, in two and higher dimensional models with sign problem, there remain significant challenges.

The advent of quantum computers and simulators \cite{PRXQuantum.2.017003} has enabled new approaches for this old problem, involving both quantum and classical computing.  Many Hamiltonians of interest naturally decompose into two pieces $H=H_A + H_B$, in which case quantum annealing is an option given experimental ability to time evolve with a linear combination of $H_A$ and $H_B$.  However, for digital noisy intermediate scale quantum (NISQ) devices \cite{Preskill2018quantumcomputingin}, with limited circuit depth, it may be more viable to use a discrete, bang-bang approach \cite{varreview, PhysRevX.7.021027}.           

In this context, bang-bang refers to a protocol which alternates between evolution with the two Hamiltonians $H_A$ and $H_B$.  One example that has attracted significant interest recently is the quantum approximate optimization algorithm (QAOA) \cite{qaoa}.  QAOA was originally proposed as a way to use quantum computers to find solutions to classical optimization problems, described by a classical cost function $H_Z$.  QAOA is a variational wave-function prepared by alternating between evolving with the Hamiltonian $H_A = H_Z$ and $H_B$ corresponding to a transverse field. The evolution times are variational parameters that are chosen to minimize the energy cost function.  The limited gate set required and efficiency as demonstrated in several numerical studies makes QAOA an appealing choice for NISQ devices.  Importantly, QAOA has been generalized to approximate (quantum) ground states of strongly correlated systems and to prepare highly-entangled states on near-term quantum devices \cite{variational,10.21468/SciPostPhys.6.3.029,https://doi.org/10.48550/arxiv.1906.08948,PRXQuantum.1.020319,PRXQuantum.2.010309,longrange,doi:10.1073/pnas.2006337117,doi:10.1073/pnas.2006373117}.

Such approaches inspired a bang-bang protocol intended for classical computers, which involves the same alternating variational ansatz but executed in imaginary, not real, time.  This ansatz, dubbed the variational imaginary time ansatz (VITA) \cite{vita}, exhibits exponential speedups compared with conventional imaginary time evolution when targeting the critical point of the one-dimensional transverse field Ising model.  VITA's initial successes suggest that it can be used to accurately represent ground states, yet it remains a puzzle how such a simple ansatz, depending only on such a small number of variational parameters, can describe the ground state so well.  Understanding the expressivity of VITA is important from both a fundamental perspective (it reveals a new way of thinking about ground states) and a practical, computational one. 

In short, QAOA-inspired and VITA approaches are promising bang-bang many-body ansatzes for quantum and classical computers, respectively. Nonetheless, the efficiency and limitations of both are not very well understood and would benefit from scalable numerical simulations. In this work, we propose the use of tensor networks in exploring both approaches (see \cite{https://doi.org/10.48550/arxiv.2206.07024} for a recent work using tensor network methods in analyzing QAOA applied to optimization problems). Tensor networks allow for an efficient classical description of many-body wave-functions provided these are moderately entangled. We use tensor networks to systematically investigate the use of VITA and QAOA to represent/prepare ground states in the infinite system limit, in contrast to previous works involving finite sizes. Our hope is that this will allow us to learn useful lessons that may also apply for more entangled scenarios, where tensor networks can no longer provide an efficient simulation of VITA and QAOA, but where the latter classical and quantum approaches might still operate effectively.

\section{Setup}

\subsection{VITA and QAOA}

Many Hamiltonians $H$ can be decomposed into two individually tractable sub-Hamiltonians $H = H_A + H_B$. For example, in a one-dimensional lattice with nearest-neighbor interactions, we can write the Hamiltonian as $H = \sum H_{i, i+1} = H_{\text{odd}} + H_{\text{even}}$ where  $H_{\text{odd}} = \sum H_{2i-1, 2i}$ and $H_{\text{even}}= \sum H_{2i, 2i+1}$. Note that terms in $H_{\text{odd}}$ ( or $H_{\text{even}}$ ) commute among themselves, making each sub-Hamiltonian relatively easy to analyze.

Starting from an initial state $\ket{\psi_0}$, we evolve with alternating Hamiltonians $H_A$ and $H_B$ in imaginary time (VITA) with $P$ pairs of variational time parameters $\{\bm{\alpha, \beta} \} =  \{\alpha_1, \beta_1; \alpha_2, \beta_2; ... ; \alpha_P, \beta_P\}$:
\begin{equation}
\label{imag}
\ket{\psi_P(\bm{\alpha, \beta})} = \frac{ \prod_{p=1}^{P} e^{-\beta_p H_B} e^{-\alpha_p H_A}  \ket{\psi_0} } {\norm{\prod_{p=1}^{P} e^{-\beta_p H_B} e^{-\alpha_p H_A}  \ket{\psi_0} }},
\end{equation}
or in real time (generalized QAOA)
\begin{equation}
\ket{\psi_P(\bm{\alpha, \beta})} =\prod_{p=1}^{P} e^{- i \beta_p H_B} e^{- i \alpha_p H_A}  \ket{\psi_0} .
\label{real}
\end{equation}
Above, the product is ordered, in that $\prod_{p=1}^P O_p$ stands for $O_P \times O_{P-1} \times \cdots \times O_2 \times O_1 $.

For a certain initial state and a given depth $P$, an optimal set of parameters $\{ \bm{\alpha, \beta} \}$ are obtained by minimizing the energy $E_P(\bm{\alpha, \beta}) = \ev{H}{\psi_P(\bm{\alpha, \beta})}$.

\subsection{MPS}

For one dimensional many-body systems, matrix product state (MPS) methods \cite{White, Fannes, itebd, itebd1, Verstraete}, have proved to be a powerful tool to represent physical wavefunctions. Consider a translation invariant state $\ket{\psi}$ of an infinite one-dimensional lattice system. An infinite, translationally invariant MPS representation of state $\ket{\psi}$ consists of a pair of tensors $ \{\Gamma, \lambda\}$ \cite{Fannes, itebd}. Here $\Gamma$ is a three-index tensor of dimension $d\times D\times D$ and $\lambda$ is a diagonal matrix of dimension $D\times D$, where $d$ is the dimension of the Hilbert space of each lattice site and $D$ is the so-called bond dimension of the MPS. The cost of a classical computation scales with $D$ as $O(D^3)$. Here we are specifically interested in states $\ket{\psi}$ that can be either exactly represented, or well approximated, by an MPS with a sufficiently small bond dimension $D$, such that we can afford a classical simulation. For instance, the largest bond dimension considered below is $D_{\max}=40$, which is sufficient for the study we wanted to conduct. Significantly larger values of $D_{\max}$ can in principle be considered, ranging from the low $1000s$ with a desktop all the way up to $100,000s$ with a supercomputer \cite{ganahl2022density}.

We represent initial state $\ket{\psi_0}$ using a set of MPS tensors $\{\Gamma_0, \lambda_0\}$ with physical dimension $d$ and bond dimension $D_0$ and then evolve it with alternating Hamiltonians $H_A$ and $H_B$ in either imaginary time or real time, as described in Eqs.~\eqref{imag}\eqref{real}.
The final state $\ket{\psi_P(\bm{\alpha, \beta})}$ is represented by tensors $\{ \Gamma_P(\bm{\alpha, \beta}), \lambda_P(\bm{\alpha, \beta}) \}$ with physical dimension $d$ and bond dimension $D_P$. 

Note that the bond dimension $D_p$ can grow exponentially with number of layers $p$ as $D_p = \kappa^{p} D_0$, where $\kappa$ is determined by the dimensions of local gates.
In practice, we set a maximum bond dimension $D_{\text{max}}$. After certain steps when $D_p$ reaches $D_{\text{max}}$, we perform truncation to the matrix product state tensors such that bond dimension stays fixed as $D_{\text{max}}$. This implies that the MPS represents the final state $\ket{\psi_P(\bm{\alpha, \beta})}$ approximately, typically by neglecting the smallest coefficients in its Schmidt decomposition. This approximation may still be accurate, if the Schmidt coefficients that are neglected are small enough. 

\section{Numerical results}

\subsection{Imaginary time evolution}

As an example, we apply the bang-bang algorithm to the infinite-size one dimensional quantum Ising model 
\begin{equation}
H = -  \sum_i Z_i Z_{i+1} - h  \sum_i X_i 
\label{Ising}
\end{equation}
and compare the performance of bang-bang algorithm with that of projector method.

The standard projector approach of obtaining the ground state of a Hamiltonian $H$ is to simulate an evolution in imaginary (euclidean) time as given by $\ket{\Psi} = \lim_{\tau \rightarrow \infty} \ket{\psi(\tau)}$, for $\ket{\psi(\tau)} \equiv \frac{e^{-\tau H} \ket{\psi(0)}}{\norm{e^{-\tau H} \ket{\psi(0)}}} $. One expects that as $\tau$ increases, $\ket{\psi(\tau)}$ will monotonically converge towards the exact ground state $\ket{\Psi}$. More specifically, as numerically checked using the iMPS algorithm, the fidelity per site increases exponentially with $\tau$.
For the non-critical (gapped) case, it is also expected that, to prepare the ground state to a certain fidelity, the total imaginary time required should scale with $1/\Delta$ where $\Delta$ is the energy gap between ground state and the first excited state. For Ising Hamiltonian in \eqref{Ising}, $\Delta = 2\abs{h-1}$. Here this statement is verified numerically by finding the scaling relation between error in fidelity per site $(1-f)$ between $\ket{\psi(\tau)}$ and the ground state $\ket{\psi_\text{g.s.}}$ and $\tau\Delta$(Fig. 2), where fidelity per site $f$ is calculated using matrix product state representation. We find that for standard projector method, $(1-f) \sim e^{- \tau \Delta}$. We conclude that for a fixed target fidelity per site and varying energy gap $\Delta$, the total required time scales as $\tau\sim 1/\Delta$.

\begin{figure}[H]
\includegraphics[width=0.43\textwidth]{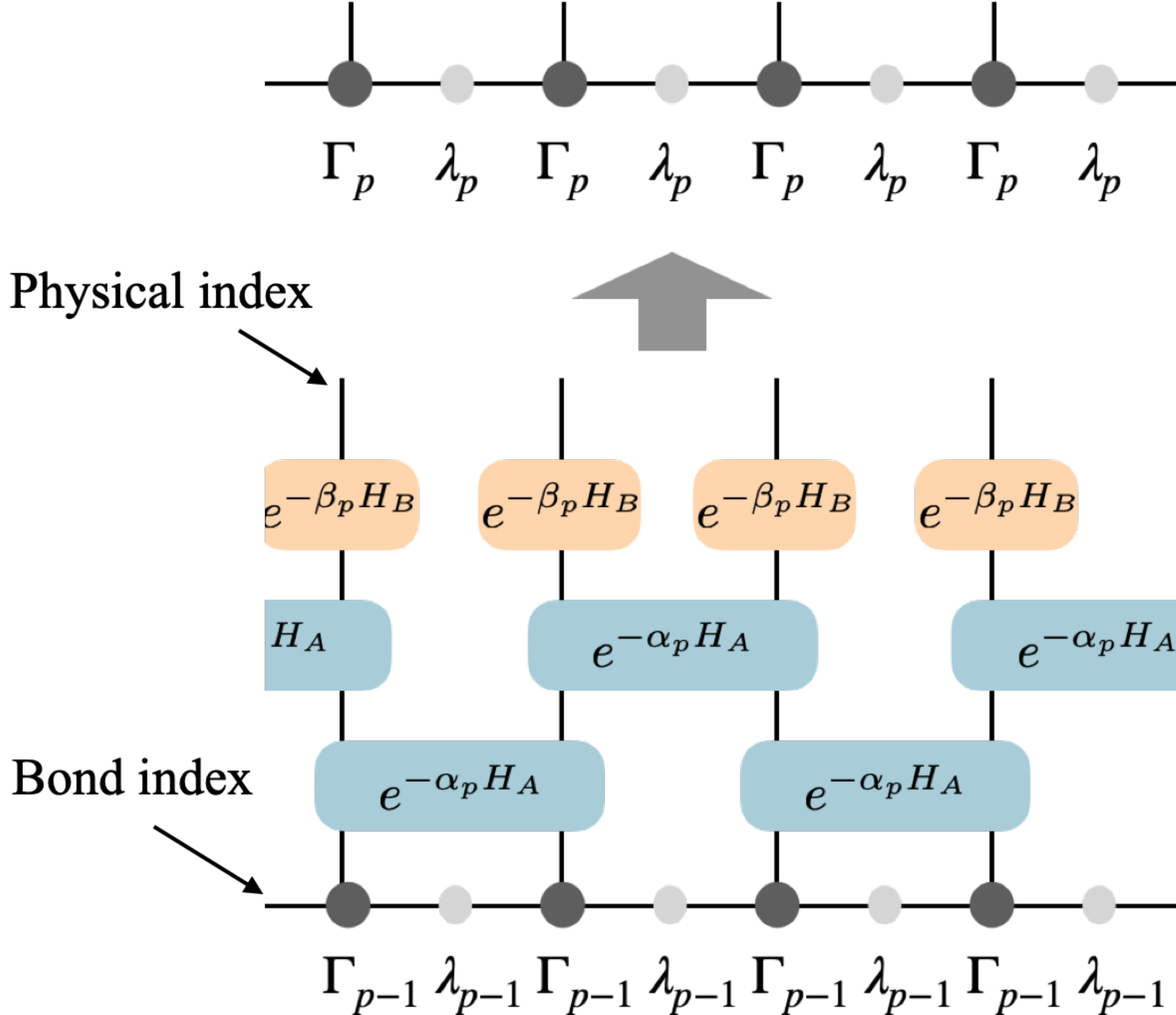}
\caption{One layer of evolution for infinite MPS with two-site translation invariance. Each step updates iMPS $\{ \Gamma_{p-1}, \lambda_{p-1} \}$ to $\{ \Gamma_p, \lambda_p \}$.}
\label{fig:setup}
\end{figure}

\textit{Imaginary time bang-bang algorithm}:
Now we describe in detail how to approximate the ground state of Ising model using VITA. In the case of Ising model, a natural way to decompose the Hamiltonian into two parts is taking $H_A = -  \sum_i Z_i Z_{i+1} $ and $ H_B = - h  \sum_i X_i $. We start from an initial product state iMPS $\{\Gamma_0, \lambda_0 \}$ and evolve with $H_A$ and $H_B$ in an alternating fashion for $P$ steps to obtain the final iMPS $\{\Gamma_P, \lambda_P \}$. For example, Fig.~\ref{fig:setup} shows the $(p+1)$-th step that brings $\{ \Gamma_p, \lambda_p \}$ to $\{ \Gamma_{p+1}, \lambda_{p+1} \}$ . 

Here since the initial state is a product state, the bond dimension of the initial iMPS is $D_0=1$ and the physical dimension is $d = 2$ for spin-1/2 systems. As $p$ increases, the bond dimension $D_p$ grows as $D_p=2^p$ before it gets to the maximal bond dimension, here set to be $D_\text{max} = 40$. Once $D$ reaches $D_\text{max}$ we truncate the iMPS and keep only the largest $D_\text{max}$ values of each bond index. For example, $D_{(p=4)} = 16$, $D_{(p=5)} = 32$, and $D_{(p\geq6)} = 40$. 

For a given $P$, the trial iMPS is a function of variational parameters $\{ \bm{\alpha, \beta} \}$. Using an optimization algorithm (e.g. black-box optimization function \textsc{bboptimize} in \textsc{Julia} ), the minima of the energy cost function $E_P(\bm{\alpha, \beta})$ together with the optimized parameters $\{ \bm{\alpha, \beta} \}$ can be found. After the optimization procedure, we get the final iMPS $\{\Gamma, \lambda\}_P$. To see how well the optimized final state approximates the ground state, we can calculate the fidelity between $\ket{\psi_P(\bm{\alpha, \beta})}$ and the ground state $\ket{\psi_\text{g.s.}}$ obtained by iTEBD.
Details on computing fidelity per site in the context of tensor network can be found in \cite{fidelity}.

\begin{figure}[H]
  \begin{subfigure}[h]{0.23\textwidth}
  \centering
  \caption{}
    \includegraphics[width=\linewidth]{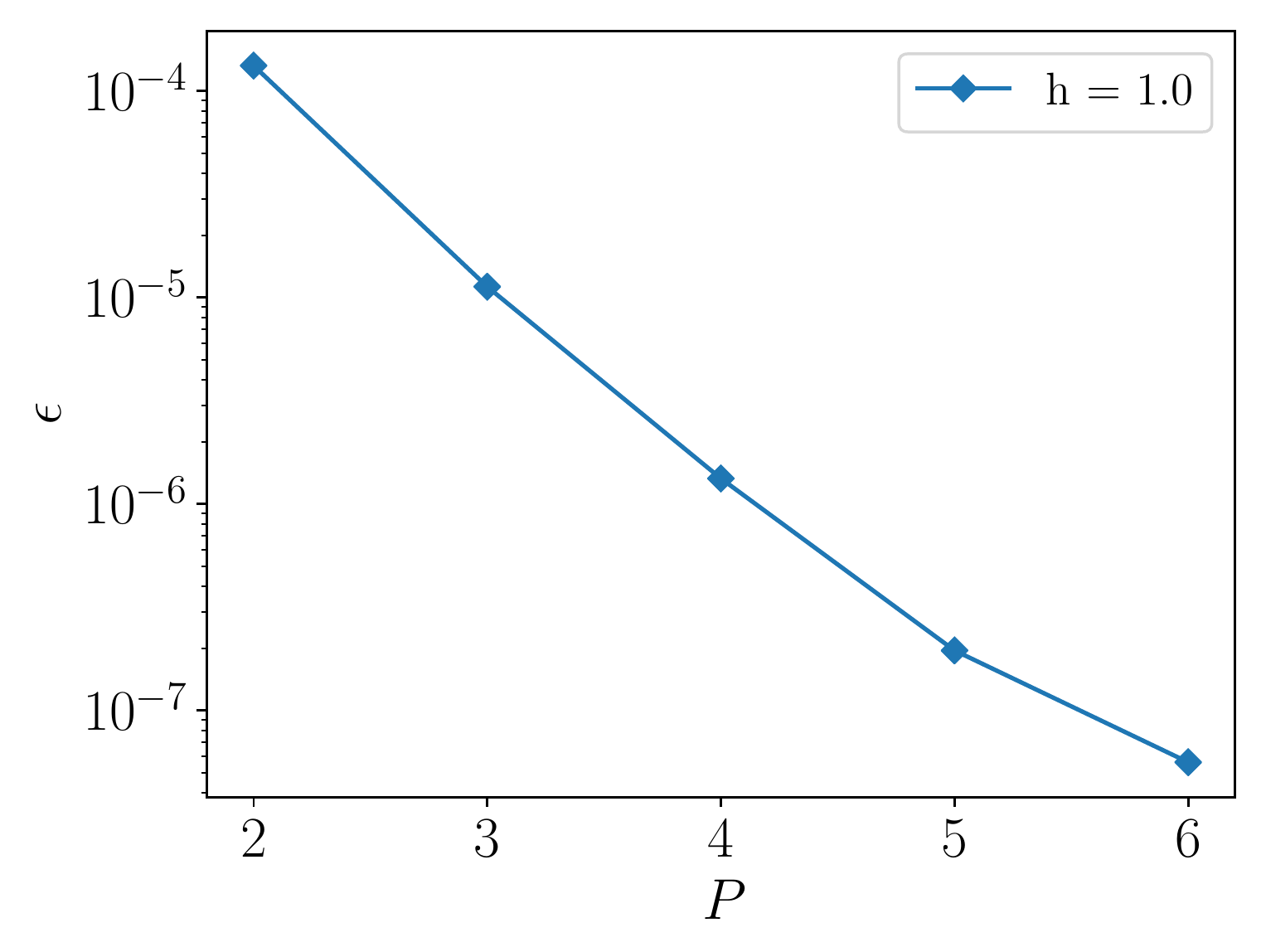}
    \label{fig:ising_energy_crit}
   \end{subfigure}
  \begin{subfigure}[h]{0.23\textwidth}
  \centering
  \caption{}
    \includegraphics[width=\linewidth]{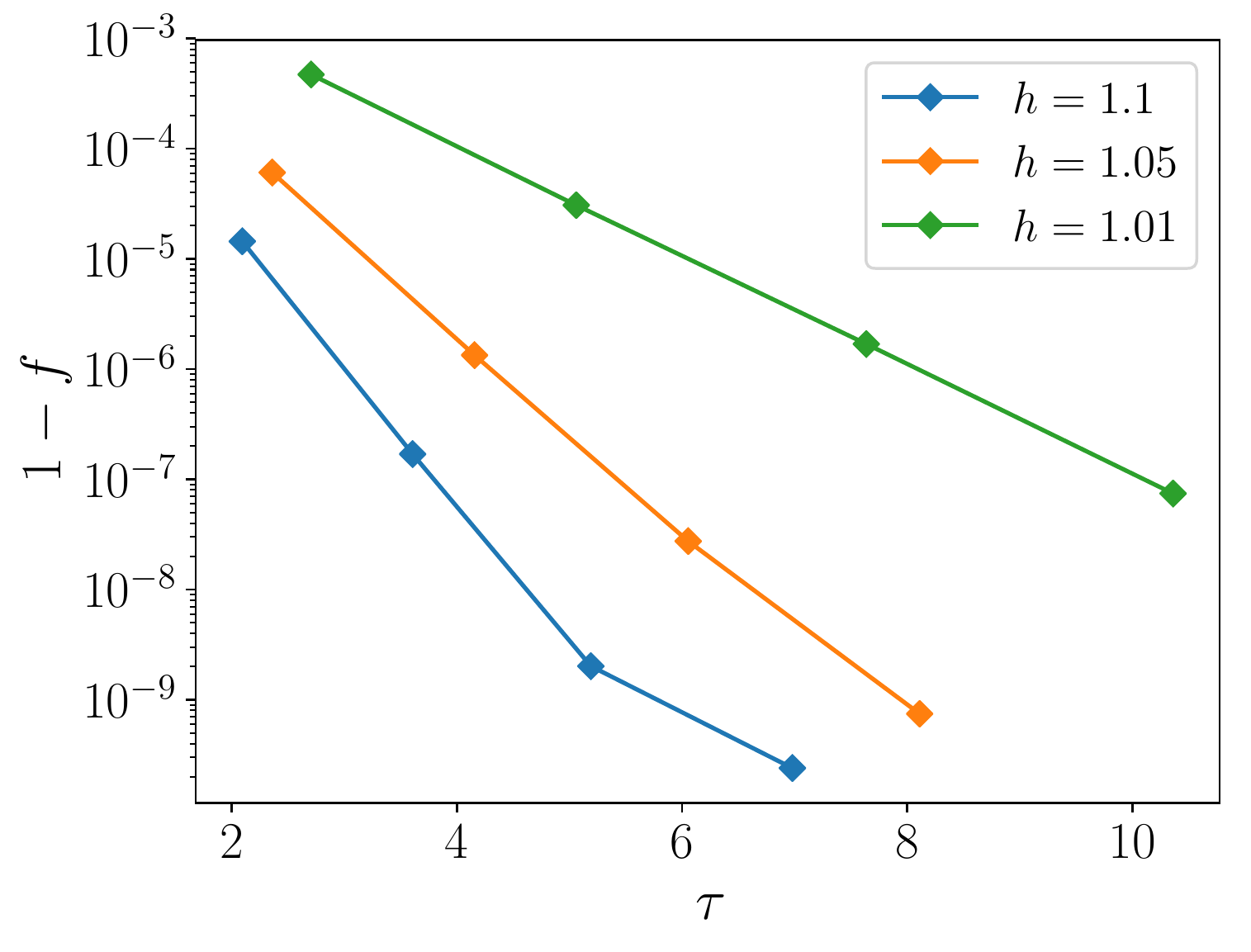}
    \label{fig:f_gapped}
   \end{subfigure}
   \\
   \begin{subfigure}[h]{0.23\textwidth}
  \centering
  \caption{}
    \includegraphics[width=\linewidth]{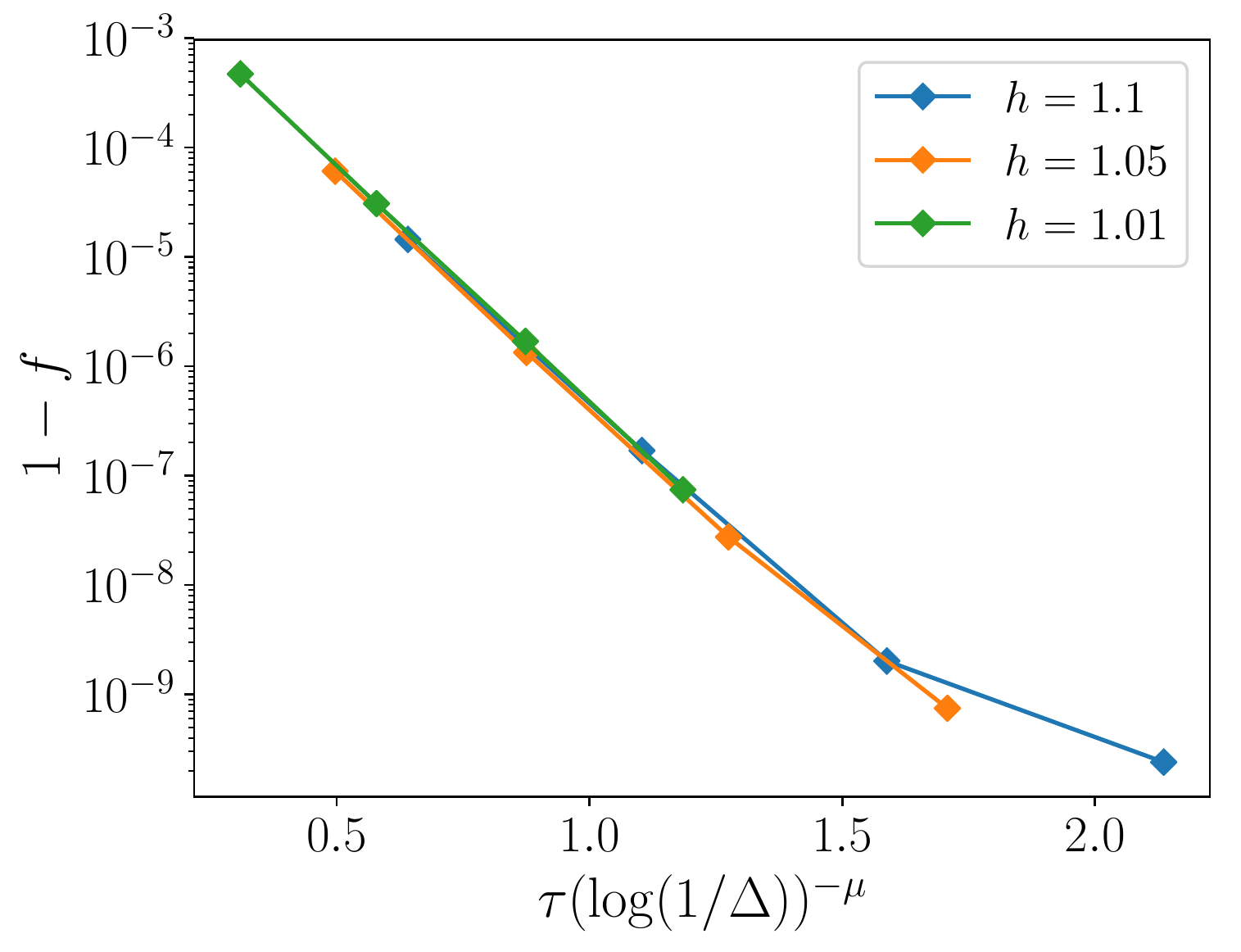}
    \label{fig:f_collapse}
   \end{subfigure}
   \begin{subfigure}[h]{0.23\textwidth}
  \centering
  \caption{}
    \includegraphics[width=\linewidth]{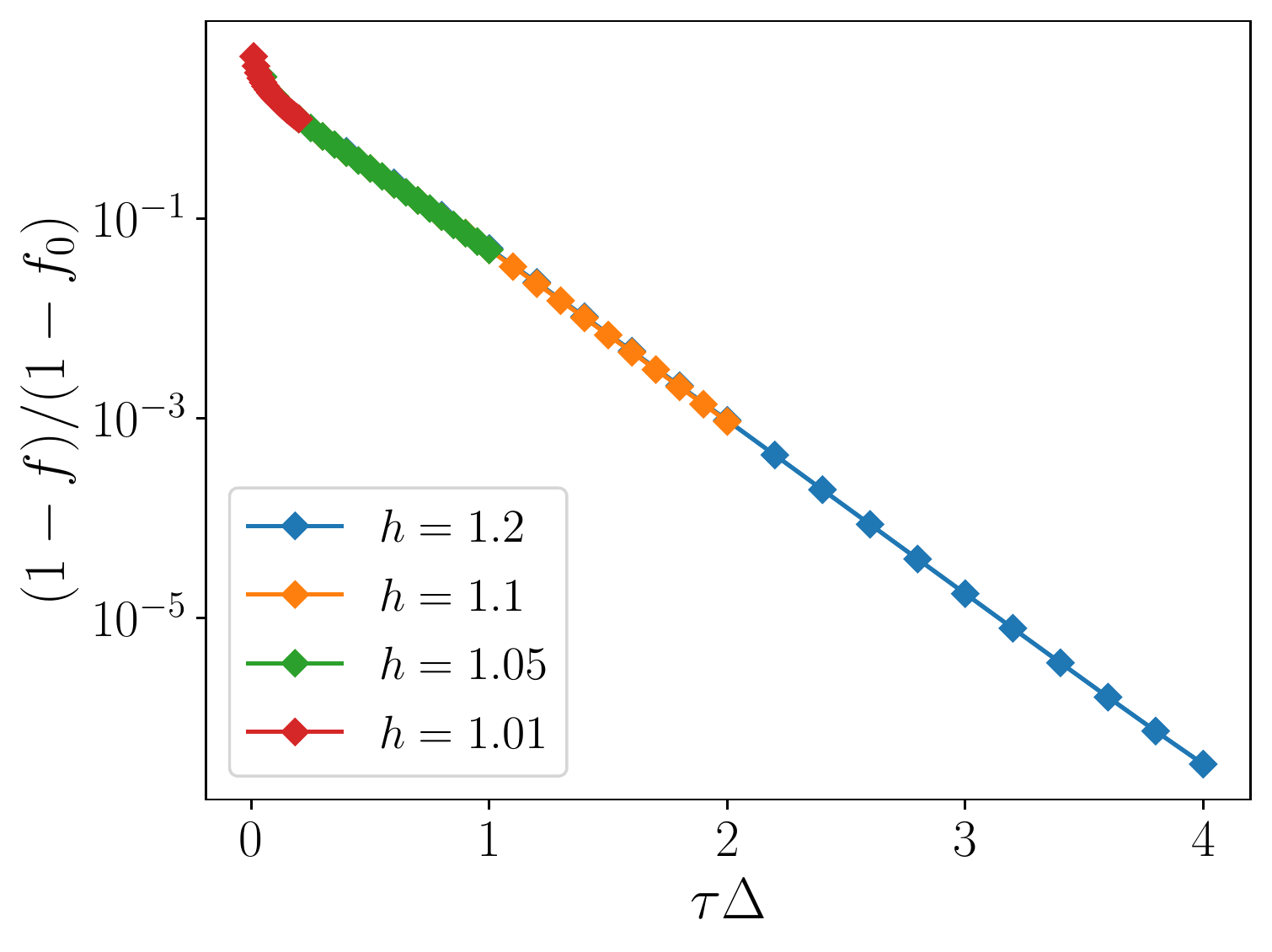}
    \label{fig:itebd_gap}
   \end{subfigure}
 \caption{Imaginary time evolution for Ising model at different values of transverse field. (a) Relative error in energy $\epsilon$ between the exact ground state energy, $E_{\text{exact}}$, and the energy of the optimized iMPS $E_P(\bm{\alpha, \beta})$ obtained by VITA. (b) Error in fidelity per site. Shown here is the case with total $P = 5$ optimization steps. (c) Collapse of error in fidelity per site $1-f = F(\tau(\log(1/\Delta))^{-\mu})$ with $\mu = 1.42$. $P = 5$. (d) Decay of error in fidelity with imaginary time, using the projector method. Given a gapped Hamiltonian, error in fidelity per site $1-f$ drops exponentially with $\tau$ when $\tau$ is large enough.}
\label{fig:ising_im}
\end{figure}

We again look at the scaling relation between the required imaginary time $\tau = \sum_{p=1}^P \alpha_p + \beta_p$ to achieve certain fidelity and the energy gap $\Delta$. The scaling behavior for the VITA ansatz is substantially different from that of imaginary time evolution by $H$ (as used in e.g. iTEBD or quantum Monte Carlo methods). Instead of $\tau \sim 1/\Delta$, here we find $\tau \sim ({\log{(1/\Delta)}})^{\mu}$, where the power $\mu$ can be determined by a scaling collapse Fig.~\ref{fig:f_collapse}. In other words, for a fixed energy gap $\Delta$, the total required time using variational optimization method is much less compared to that of standard projector methods, see Fig.~\ref{fig:itebd_gap}. This dramatic speedup was also noted for finite size simulations of VITA \cite{vita}.

\begin{figure}[H]
\includegraphics[width=0.45\textwidth]{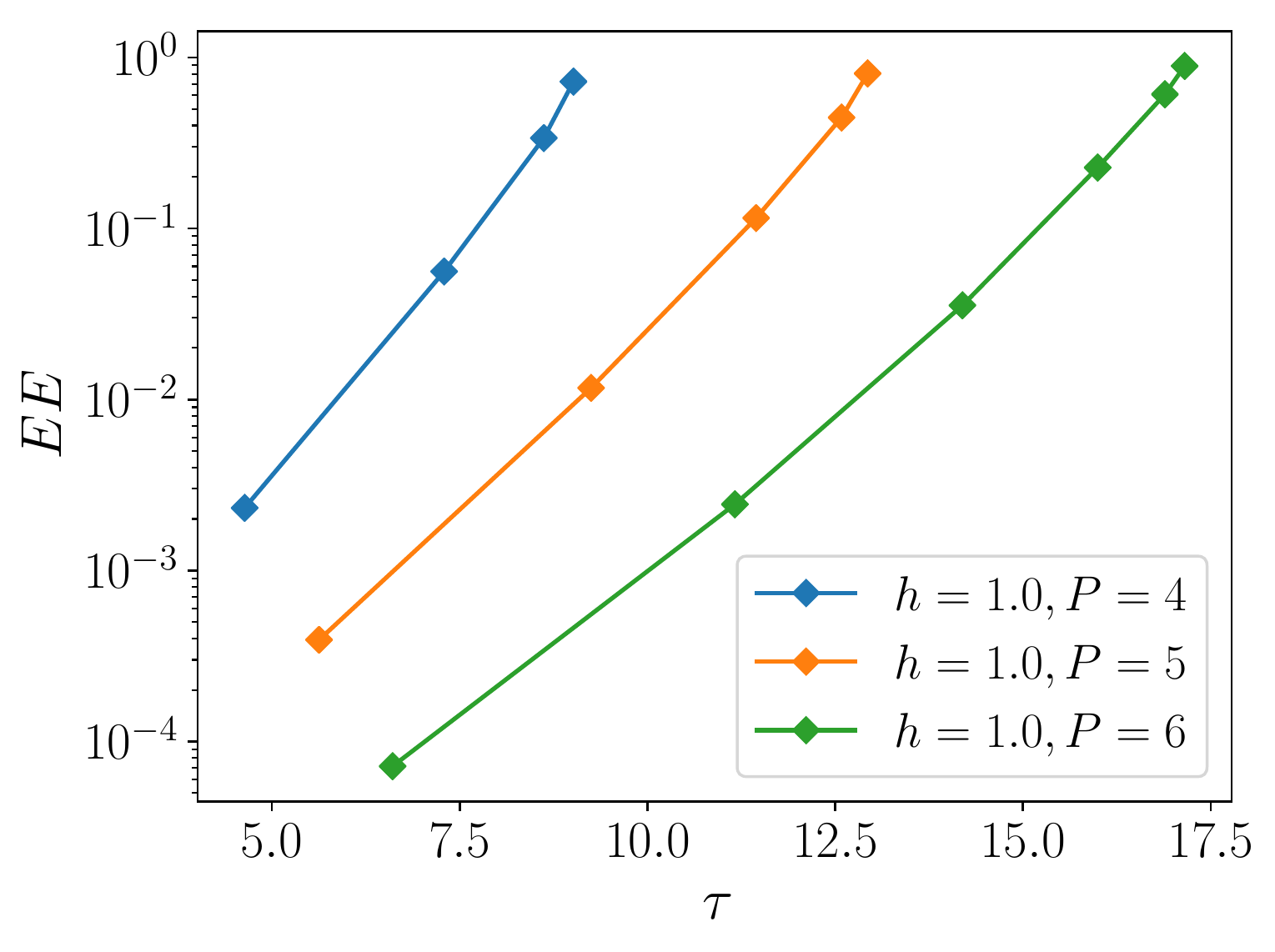}
\caption{At criticality, entanglement entropy of half infinite spin chain increases exponentially with imaginary time $\tau$.} 
\label{fig:ising_im_ee}
\end{figure}

Furthermore, information on entanglement dynamics can be obtained by calculating the entanglement entropies of half infinite chain at intermediate steps during one optimization process with fixed total number of rounds. Shown in Fig.~\ref{fig:ising_im_ee}, in the imaginary time evolution, the entanglement entropy can grow exponentially with imaginary time for the first several rounds, as the more entangled the states are, the more entanglement can be introduced into the system in one evolution step.  This was also noticed and discussed for finite size simulations of VITA \cite{vita}. 

In the appendices, we show additional results for VITA applied to the transverse field Ising model, including the accuracy of two-point correlation functions achieved by the ansatz and results for a different decomposition of the Hamiltonian into odd and even blocks.  Furthermore, as an example of the performance of VITA applied to a different model, which highlights the generality of the method, we show in the appendix results for the Blume--Capel model, which realizes a tricritical Ising universality class.

\subsection{Real time bang-bang algorithm}

Similar tensor network simulations can be used to study real time bang-bang protocols for the Ising Hamiltonian. In the real time case, the ground state is approximated as in \eqref{real} with $H_A$ and $H_B$ the same as above. While imaginary time evolution serves as an efficient computational method for finding the ground states of many body systems on classical computers, studying the real time evolution is important for quantum simulations on quantum hardware.   

\begin{figure}[H]
  \begin{subfigure}[h]{0.23\textwidth}
  \centering
  \caption{}
    \includegraphics[width=\linewidth]{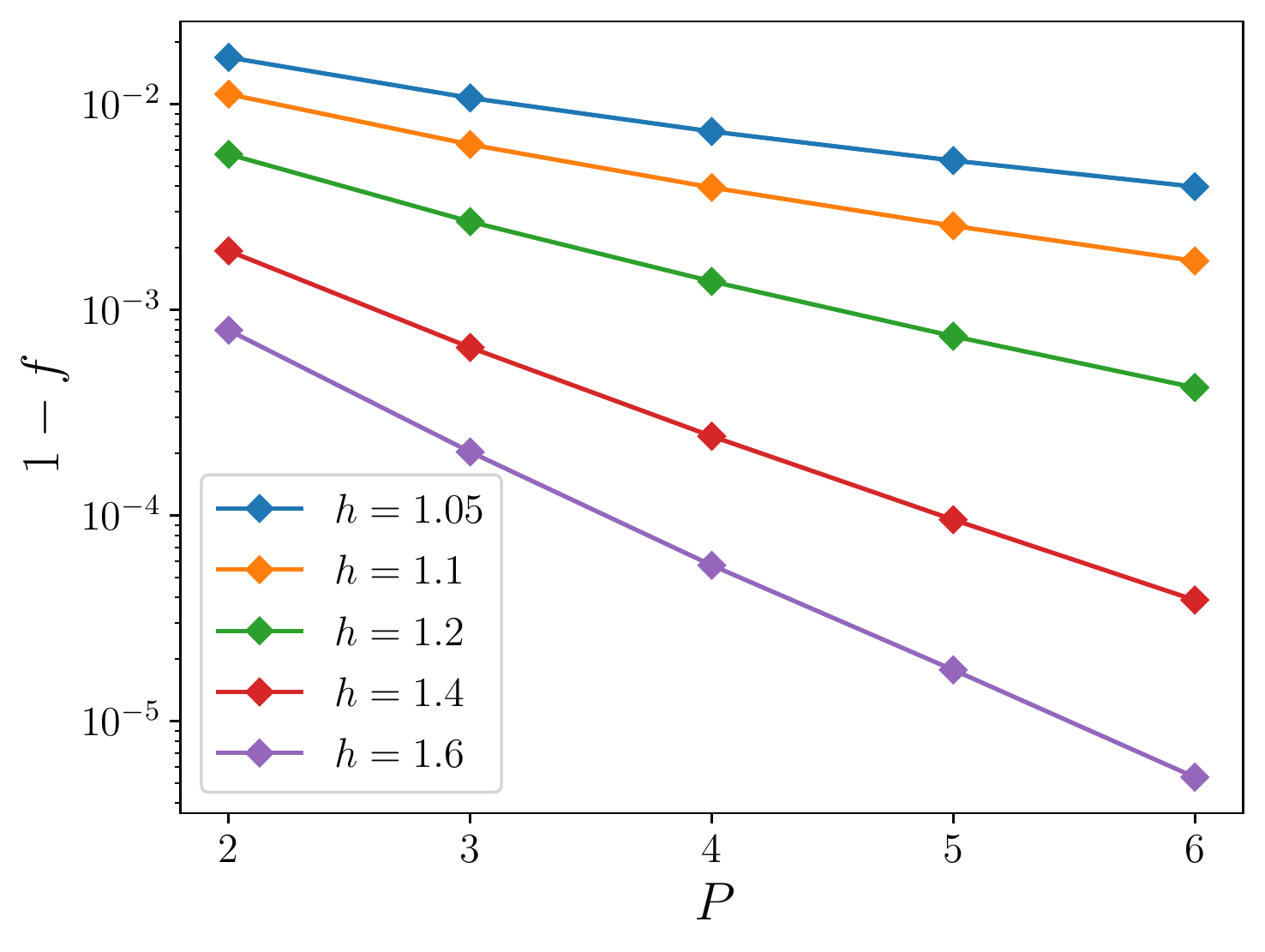}
    \label{fig:real_P}
   \end{subfigure}
  \begin{subfigure}[h]{0.23\textwidth}
  \centering
  \caption{}
    \includegraphics[width=\linewidth]{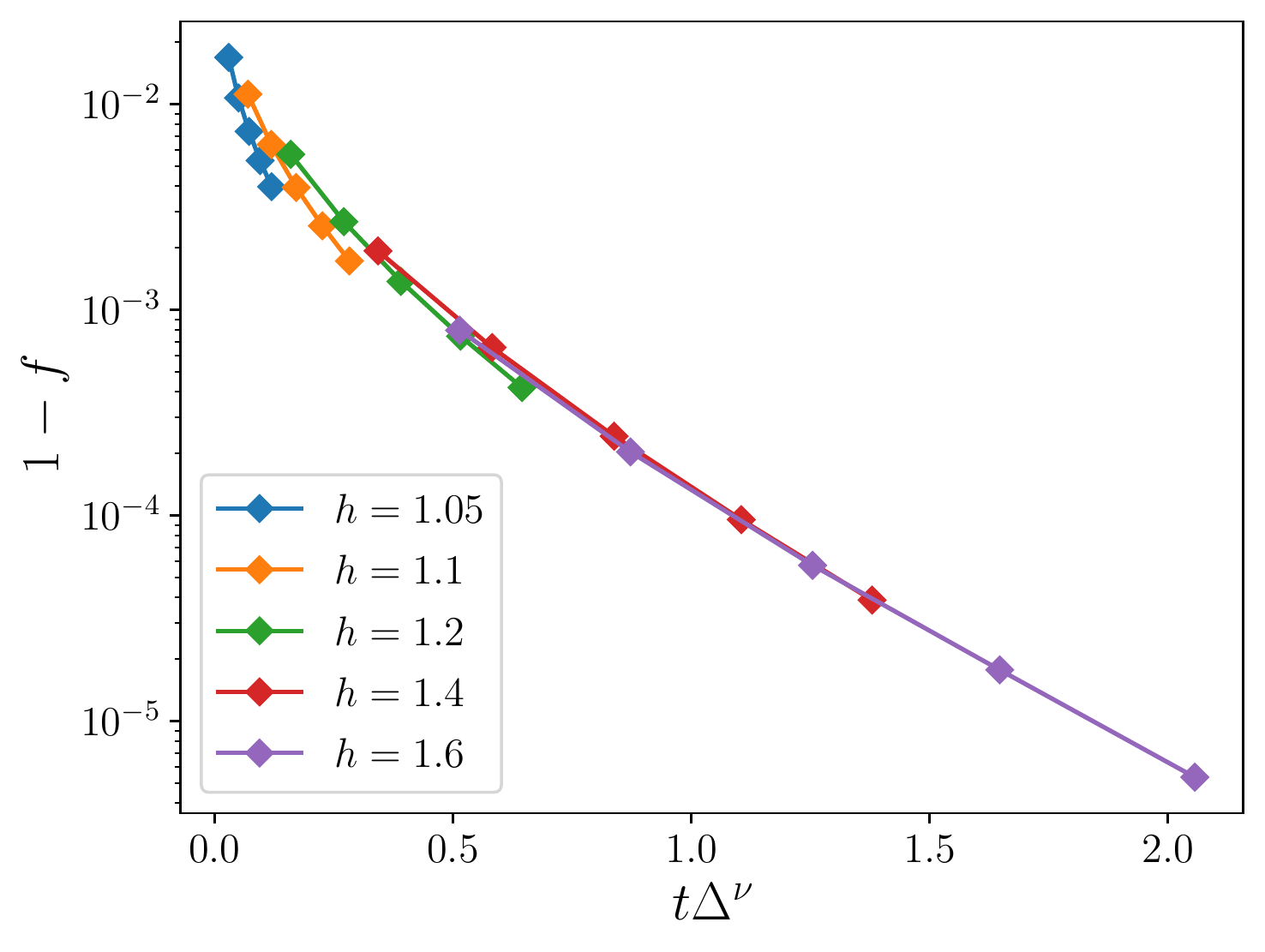}
    \label{fig:real_t}
   \end{subfigure}
   \caption{Variational real time evolution. (a) Error in fidelity per site decays exponentially with number of real time parameters. (b) Scaling collapse supporting $1-f = F(t \Delta^{\nu})$, where $t$ is total time used in ansatz, $\Delta$ is energy gap, and $\nu = 1.3$.}
\label{fig:ising_real}
\end{figure}

We compare the optimized $\ket{\psi_P}$ from the real-time bang-bang algorithm with the ground state $\ket{\psi_\text{g.s.}}$. We find that in the gapped regime of the model, the error in fidelity per site drops exponentially with the number of applied steps $P$ (Fig.~\ref{fig:real_P}). Thus, away from the critical point, only a small number of variational steps can already approximate the ground state to high accuracy.  However, the scaling of the fidelity error with the energy gap is very different from the imaginary time case.  As the circuit is built from unitary gates, Lieb--Robinson bounds limit the growth of correlations, and we find the scaling $1-f = F(t \Delta^{\nu})$, with $\nu=1.3$ (Fig.~\ref{fig:real_t}).

\section{Discussion}
We use matrix product states (MPS) to explore the performance of VITA and generalized QAOA to represent or prepare ground states of one-dimensional infinite quantum spin chains. 
There are three key findings in our study.

First, VITA approximates a gapped ground state of 1d quantum Ising model with total imaginary time $\tau$ scaling as $\log(1/\Delta)$ in contrast to standard projector methods where $\tau \sim 1/\Delta$. One reason for this speed-up is that evolving alternatively by $H_A$ and $H_B$, which have gaps of $\mathcal{O}(1)$, may allow for much faster progression in Hilbert space than evolving by the original Hamiltonian, whose gap $\Delta$ might be much smaller. It would be interesting to understand the universality of the scaling between the gap and the imaginary time. 

Second, during the alternating imaginary time evolution, entanglement entropy grows exponentially with imaginary time $\tau$, which is a feature not possible in local real time evolution. 

Third, in the real time evolution algorithm, we find that the accuracy for preparing a gapped ground state in 1d quantumn Ising model improves exponentially with number of alternating steps, demonstrating its efficiency for near-term quantum platforms.  While similar results for VITA were obtained previously for finite systems, here we demonstrate the utility of tensor network implementations by verifying these scaling claims in the {\it infinite} system limit.

While the number of imaginary time evolution steps needed for a target accuracy is dramatically fewer than standard time evolution methods such as TEBD, additional computational costs arise from the optimization part and can be expensive when the number of steps is large. Better optimization techniques may help reduce the computational cost and make this bang-bang algorithm more efficient in practice. 

Though here we only test the efficiency of bang-bang algorithm in one-dimensional systems with matrix product states, we expect that this algorithm should also work for higher dimensional systems in which other tensor network methods such as projected entangled pair states (PEPS) can apply \cite{Verstraete}. To probe the efficacy of VITA in these larger contexts, it will be important to explore the tradeoff between cost of optimization and reduced number of imaginary time steps.

{\it Acknowledgements:}
R.W. acknowledges support from the Perimeter Institute for Theoretical Physics through the Visiting Graduate Fellowship Program where the major part of this research was done. G.V. is a CIFAR fellow in the Quantum Information Science Program, a Distinguished Invited Professor at the Institute of Photonic Sciences (ICFO), and a Distinguished Visiting Research Chair at Perimeter Institute. Research at Perimeter Institute is supported in part by the Government of Canada through the Department of Innovation, Science and Economic Development Canada and by the Province of Ontario through the Ministry of Colleges and Universities.

\bibliography{ref}

\appendix

\section{Tricritical Ising model}

We take the Blume-Capel model as one realization of the tricritical Ising model and explore the performance of the bang-bang VITA algorithm in this case. At the tricritical point, the Hamiltonian can be written as
\begin{equation}
    H = - \frac{1}{\zeta} \sum_{n=1}^{N} \{ S^x(n) S^x(n+1) - \alpha (S^x(n))^2 - \beta S^z(n) \}
\end{equation}
and can be decomposed into two parts $H_x =  - \frac{1}{\zeta} \sum_{n=1}^{N} \{ S^x(n) S^x(n+1) - \alpha (S^x(n))^2 \}$ and $H_z = - \frac{1}{\zeta} \sum_{n=1}^{N} \{ - \beta S^z(n) \}$. The numerical values \cite{Henkel} for the critical point are $\alpha = 0.910207(4)$, $\beta = 0.415685(6)$, $\zeta = 0.56557(50)$. We choose the initial state $\ket{\psi_0}$ to be the ground state of $H_x$ and evolve the state with $H_z$ and $H_x$ alternatively
\begin{equation}
\ket{\psi_P(\bm{\alpha, \beta})} =\prod_{p=1}^{P} e^{- \beta_p H_z} e^{- \alpha_p H_x}  \ket{\psi_0} .
\end{equation}

\begin{figure}[H]
\centering
\includegraphics[width=0.48\textwidth]{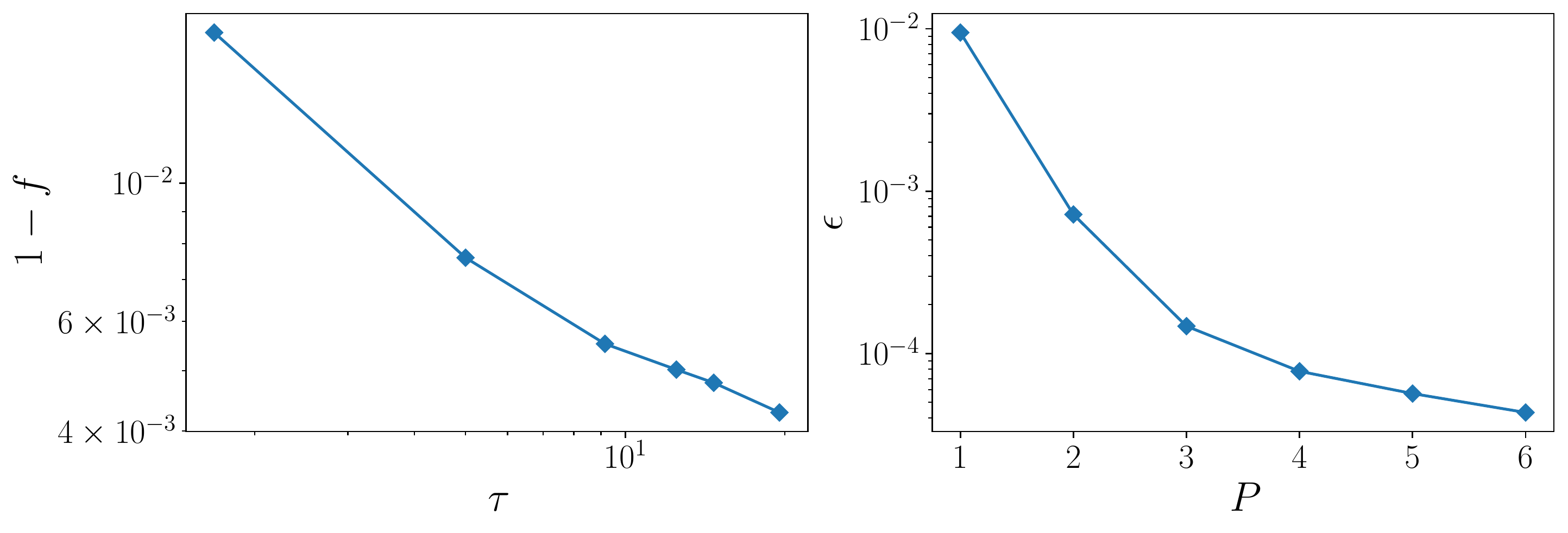}
\caption{Error in fidelity per site $1-f$ as a function of imaginary time $\tau$; relative error in energy $\epsilon$ as a function of total optimization steps $P$.} 
\label{fig:tri}
\end{figure}

\begin{figure}[H]
    \centering
    \includegraphics[width=0.4\textwidth]{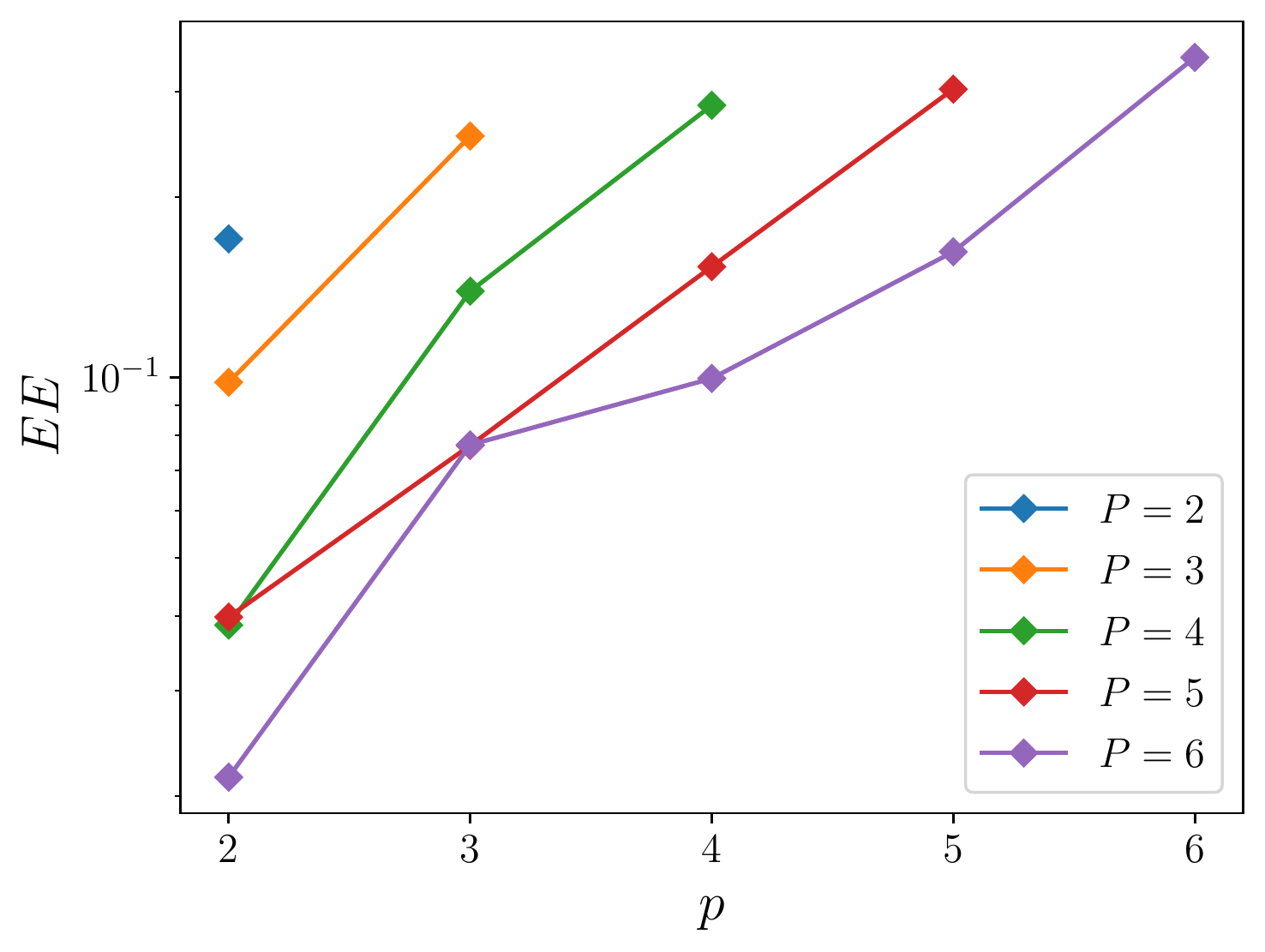}
    \caption{Entanglement growth in tricritical Ising spin chain as a function of depth $p$ during bang-bang evolution. }
    \label{fig:EE_tri}
\end{figure}

The results for fidelity and energy error are shown in Fig.~\ref{fig:tri}, and the results for bipartite entanglement growth are shown in Fig.~\ref{fig:EE_tri}.  Note that for higher $p$, the truncation error due the maximum bond dimension cutoff $D_{max}=81$ must be taken into account; the results in such cases do not purely reflect the performance of the VITA ansatz in itself. 

\section{VITA two-point correlator}

At criticality, we can also compute the two-point correlators to check if the optimized iMPS obtained by VITA gives a good approximation of the ground state. We apply the same imaginary time procedure to the infinite-size 1d TFIM as in the main text and plot the two-point correlator as a function of separation $r$ between spin operators, 
\begin{equation}
    C(r) = \ev{X_i X_{i+r}} - \ev{X_i}\ev{X_{i+r}}.
\end{equation}

\begin{figure}[H]
\centering
\includegraphics[width=0.4\textwidth]{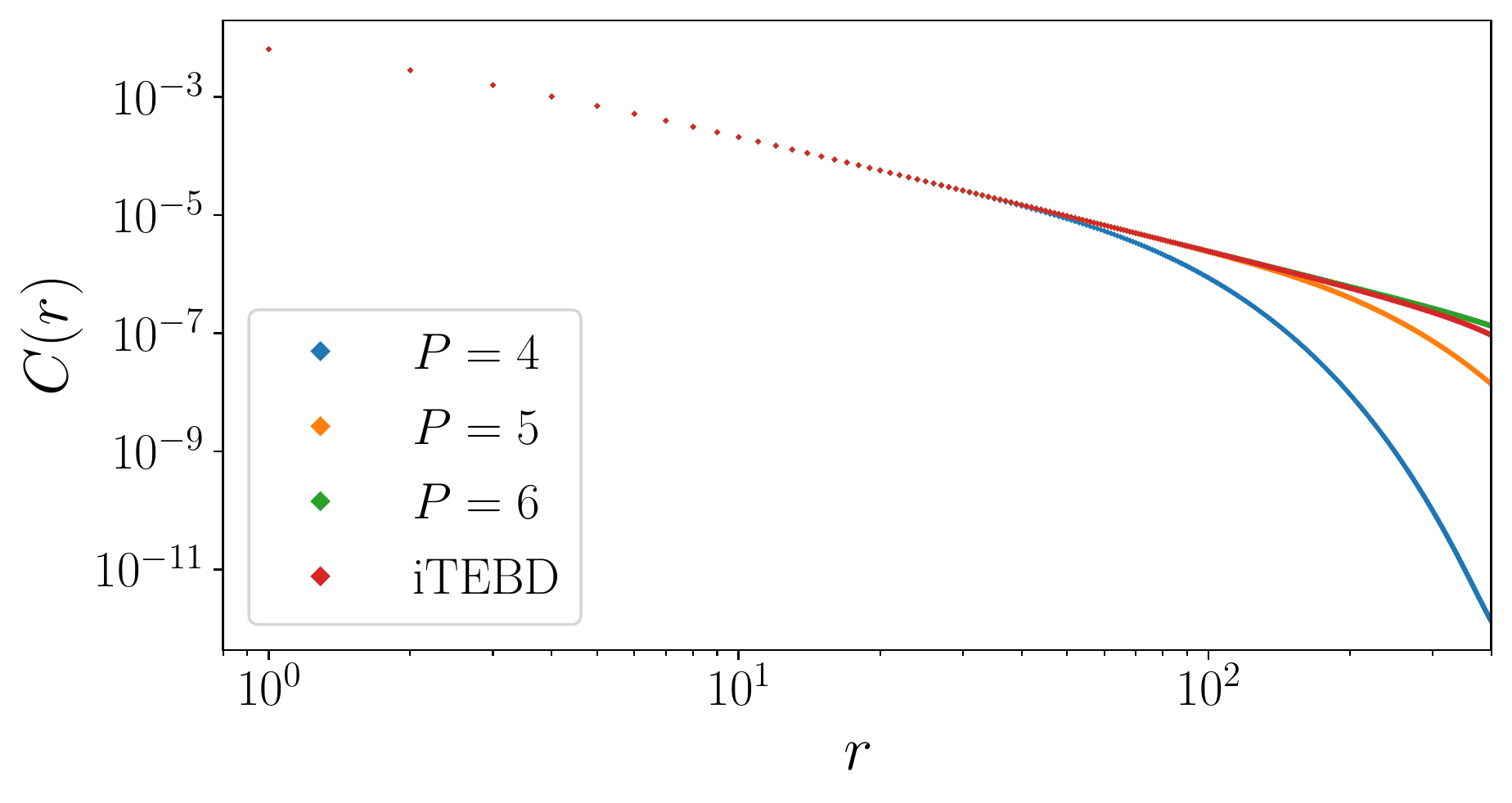}
\caption{Two-point correlator for Ising model at criticality in log-log scale plot. For the iTEBD case, bond dimension $D = 40$; for VITA cases with total steps $P = 4, 5, 6$, bond dimensions of the final states are correspondingly $D = 16, 32, 40$.  }
\label{fig:corr}
\end{figure}

As can be seen from Fig. ~\ref{fig:corr}, as the number of total steps $P$ increases, two-point correlation function gets closer and closer to a linear decay in the log-log scale (equivalently power-law decay in the normal scale), suggesting that the iMPS obtained by this variational approach indeed captures the ground states well.

\section{VITA with different splittings}

Given a Hamiltonian $H$, there could be multiple ways of decomposing the Hamiltonian into two sub-Hamiltonians, $H=H_1 + H_2$. Here we revisit the performance of VITA in the case of TFIM using a different choice of decomposition, $H = H_{\text{odd}} + H_{\text{even}}$ where $H_{\text{odd}} = \sum H_{2i-1, 2i}$ and $H_{\text{even}}= \sum H_{2i, 2i+1}$.

We show that for this odd-site and even-site decomposition, VITA again approximates the gapped ground state of 1d TFIM with total imaginary time $\tau \sim ({\log{(1/\Delta)}})^{\mu}$, but with a different exponent $\mu$, compared to Fig.~\ref{fig:f_gapped}. 

\begin{figure}[H]
\centering
\includegraphics[width=0.4\textwidth]{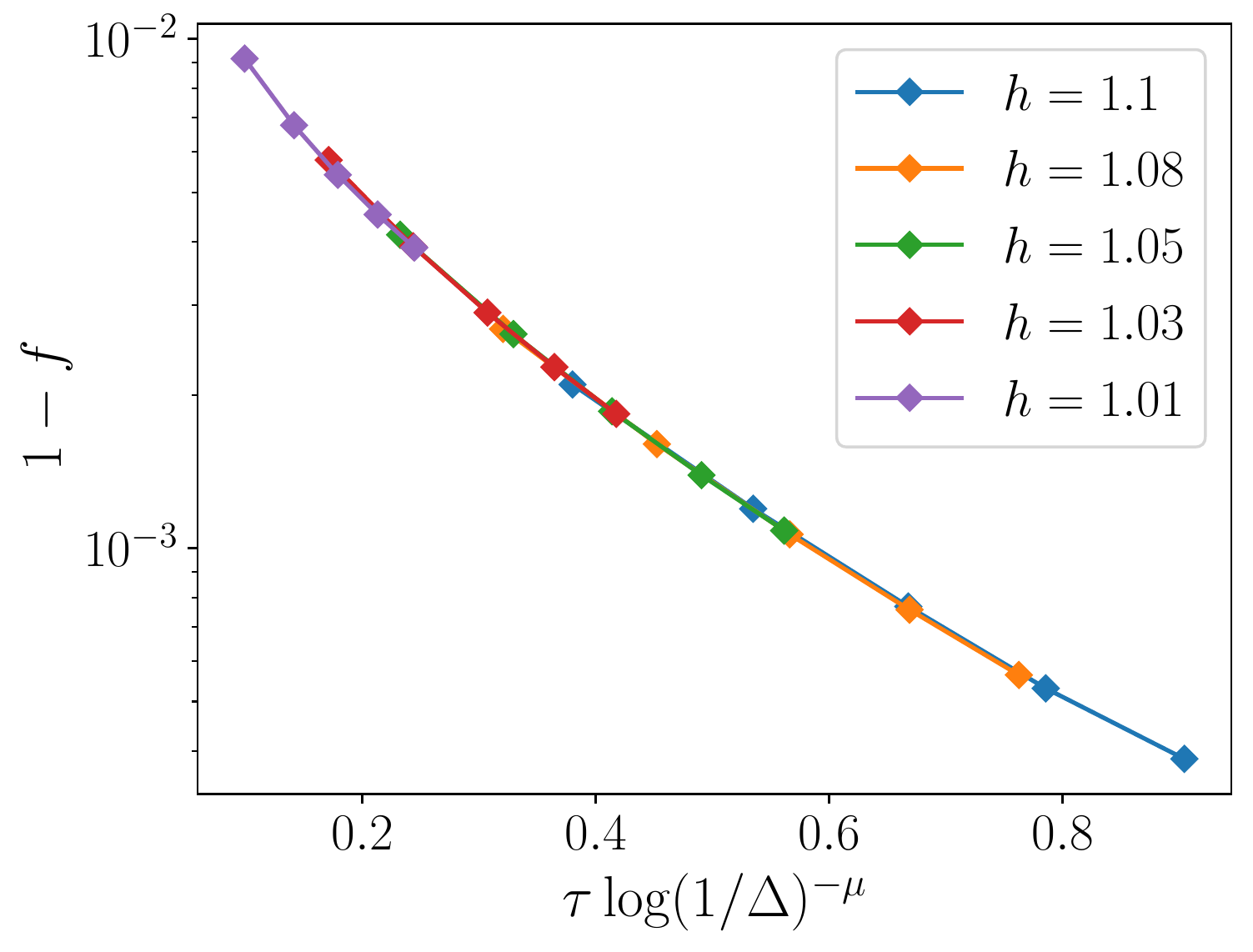}
\caption{Odd-site and even-site splitting. $H = H_{2i-1,2i} + H_{2i, 2i+1}$. $\mu = 2.1$.}
\label{fig:AB}
\end{figure}

\end{document}